\documentclass[runningheads]{llncs}
\usepackage[T1]{fontenc}
\usepackage{hyperref}
\usepackage{subcaption}
\usepackage{amsmath}
\usepackage{graphicx}
\begin{document}
\title{Validating GWAS Findings through Reverse Engineering of Contingency Tables}

%
\author{Yuzhou Jiang\inst{1}\orcidID{0000-0001-9518-8103} \and
Erman Ayday\inst{1}\orcidID{0000-0003-3383-1081}
}

\institute{
Case Western Reserve University, Cleveland, Ohio, USA \\
\email{\{yxj466,exa208\}@case.edu}
}
\maketitle              
\begin{abstract}

Reproducibility in genome-wide association studies (GWAS) is crucial for ensuring reliable genomic research outcomes. However, limited access to original genomic datasets (mainly due to privacy concerns) prevents researchers from reproducing experiments to validate results. 
In this paper, we propose a novel method for GWAS reproducibility validation that detects unintentional errors without the need for dataset sharing. Our approach leverages p-values from GWAS outcome reports to estimate contingency tables for each single nucleotide polymorphism (SNP) and calculates the Hamming distance between the minor allele frequencies (MAFs) derived from these contingency tables and publicly available phenotype-specific MAF data. By comparing the average Hamming distance, we validate results that fall within a trusted threshold as reliable, while flagging those that exceed the threshold for further inspection. This approach not only allows researchers to validate the correctness of GWAS findings of other researchers, but it also provides a self-check step for the researchers before they publish their findings.
We evaluate our approach using three real-life SNP datasets from OpenSNP, showing its ability to detect unintentional errors effectively, even when small errors occur, such as 1\% of SNPs being reported incorrectly.
This novel validation technique offers a promising solution to the GWAS reproducibility challenge, balancing the need for rigorous validation with the imperative of protecting sensitive genomic data, thereby enhancing trust and accuracy in genetic research.

\end{abstract}

\vspace{-7mm}

\section{Introduction}
\label{section:introduction}
\vspace{-1mm}
Genome-wide association studies (GWAS) have been popular in genomic research in recent years, enabling scientists to uncover associations between genetic variants and traits or diseases in understanding disease etiology~\cite{cantor2010prioritizing,naveed2015privacy,korte2013advantages,abdellaoui202315}. 
The increasing reliance on GWAS findings introduces a critical challenge: \textbf{the issue of reproducibility}.
In the context of GWAS, reproducibility involves using the original genomic dataset and applying the same analytical approach to verify the reported outcomes~\cite{begley2015reproducibility,national2019reproducibility}. 
It plays a critical role in securing funding for GWAS projects~\cite{lin2018reproducibility}. Independent replication of research outcomes is often an expectation from grant reviewers, as it is a key indicator of scientific rigor and reliability. Without the ability to validate findings through reproducibility, securing grants for GWAS research becomes increasingly difficult. 

Verifying GWAS outcomes typically requires access to the original dataset, but privacy concerns prevent researchers from sharing sensitive genomic data due to its vulnerability to reidentification and membership inference attacks~\cite{homer2008resolving,sankararaman2009genomic,deznabi2017inference,gymrek2013identifying}. Without the data, other researchers cannot replicate the study and validate the outcomes, making independent reproducibility check impractical.



Existing privacy-preserving techniques~\cite{akgun2015privacy,dwork2006differential,jiang2022reproducibility,halimi2021privacy} for reproducibility rely on sharing additional, privacy-protected information from researchers. Although these methods use differential privacy, researchers remain concerned about potential data misuse and the risk that future attacks might compromise privacy protections. Additionally, these methods often add excessive noise, limiting the data usefulness to validation alone and restricting its application for other analyses. This highlights the need for a reproducibility validation method that avoids sharing any additional data.


In this paper, we propose a novel method that relies solely on public information and research outcomes that are allowed to be shared based on data sharing policies of institutions (e.g., the NIH Genomic Data Sharing Policy~\cite{nih_gds}) while maintaining the reproducibility of GWAS outcomes. Our approach leverages shared p-values of the Single Nucleotide Polymorphisms (SNPs) as the outcome of GWAS studies. By utilizing public reference datasets, we generate the control row in the contingency table for each SNP. We then apply a grid search to find the best-matching case row that reproduces p-values as close as possible to those in the original research findings. 
We calculate the Minor Allele Frequencies (MAFs) based on these contingency tables and compare these MAFs with publicly available MAF values associated with the same phenotype or disease. The average Hamming distance between the calculated MAFs and the public MAF values is then used for the detection.
If the average distance exceeds a predefined threshold, we flag the research findings for potential errors, prompting further investigation. Otherwise, the other researchers can trust the reported results and use them to inform their own research. Note that this approach not only helps researchers validate the accuracy of GWAS findings from other studies, but it also offers a built-in mechanism for researchers to double-check their own results before publication. In the remainder of the paper, we consider a researcher-verifier scenario while introducing our proposed scheme.

It is important to note that our approach is designed to detect unintentional errors occurring at various stages, including data collection, preprocessing, and GWAS analysis. However, it does not address scenarios where a researcher deliberately fabricates data. In such cases, ethical consequences, including damage to credibility and repercussions from funding agencies, act as strong deterrents against falsifying results.

We evaluate our proposed scheme on three genomic datasets from the OpenSNP project~\cite{opensnp}. The experimental results demonstrate that our scheme effectively identifies errors in GWAS outcomes with high confidence, even in cases where minor errors occur, such as when 1\% of SNPs are reported incorrectly. Throughout the experiments, our approach consistently performs well in real-world scenarios without requiring additional information from the original research, proving its robustness and practicality for reproducibility validation in genomic studies.


\vspace{-3mm}
\section{System and Threat Model}
\label{section:system_threat}
\vspace{-2mm}
In this section, we introduce our system and threat models. 

\vspace{-3mm}
\subsection{System Model}
\vspace{-1mm}
We consider two main entities: Alice, a researcher who conducts the research and publishes the GWAS findings, and Bob, another researcher who seeks to validate these GWAS outcomes.

Alice (Researcher): Alice conducts genome-wide association studies (GWAS) on a local genomic dataset to identify associations between specific SNPs and phenotypic traits or diseases. Due to privacy concerns, Alice cannot share the raw genomic dataset used in her research. Instead, she publishes only the results of the GWAS, specifically: 1) the SNPs that show significant associations with traits or diseases, 2) the p-values associated with the SNPs, 3) the number of samples in the case and control groups, respectively. We also assume that Alice uses a publicly accessible reference dataset as the control group data for the GWAS. Alice aims to allow external validation of her findings while maintaining the privacy of the genomic dataset, thus avoiding the sharing of any additional data beyond reported findings. 

Bob (Other Researcher/Verifier): Bob, the verifier, is another researcher interested in validating the GWAS results published by Alice. 
Bob uses public genomic datasets and Alice’s reported p-values to perform his validation. More specifically, Bob first constructs the contingency table of the control group based on publicly available SNP data, then reverse-engineers the case group row $T_s^i$ in the contingency table for each SNP $i$ by utilizing the reported p-values from Alice. Using a grid search approach, he identifies $T_s^i$ that best reproduces the p-values reported by Alice, ensuring the results align with the published findings.

Next, using the constructed contingency tables, Bob computes the Minor Allele Frequencies (MAFs) for the case group and compares them to MAF values from public phenotype-specific datasets. If the average distance between the calculated and public MAFs is below a set threshold, Bob deems Alice's findings reliable. Otherwise, Bob may request further investigation or additional information for validation, subject to Institutional Review Board (IRB) approval.

As discussed before, this mechanism can also be used by Alice as a ``self-check'', allowing Alice to perform validation on her own GWAS findings. However, in this paper, we focus on the researcher-verifier setting while introducing our scheme. The self-check scenario will be discussed further in Section~\ref{discussion}.

\vspace{-2mm}

\subsection{Threat Model}
\vspace{-1mm}
In our framework, we assume that Alice, the researcher, is honest but may unintentionally introduce errors throughout various stages of the GWAS process. These errors could occur during data collection, preprocessing, GWAS experiments, or result publication. Although Alice does not intend to fabricate data, computational mistakes or mismanagement during any stage of the workflow could lead to incorrect GWAS outcomes, potentially misleading other researchers. Our approach is designed to detect such unintentional errors and flag the findings for further investigation if inconsistencies arise during the validation process.

We also account for the possibility of multiple errors occurring independently throughout the GWAS process. Although errors may arise at different stages (e.g., during data collection and preprocessing), they are treated as independent events. Importantly, our approach is still capable of detecting such compounded errors. This is because multiple independent errors are highly unlikely to combine in a way that produces results resembling those from a correctly executed GWAS. In other words, the probability that a sequence of independent mistakes yields the same outcome as a valid process is extremely low, making it easier for our method to identify discrepancies.

It is important to note that our method does not extend to scenarios where Alice deliberately fabricates results. Intentional data manipulation, such as the creation of synthetic datasets to produce false outcomes, is beyond the scope of our work. Detecting such malicious behavior is nearly impossible in most data analysis contexts, including GWAS, without direct access to the original dataset. In these cases, verification of the findings would require examination of the raw data. Furthermore, the ethical consequences of such misconduct—ranging from reputational damage to the potential loss of funding—serve as a strong constraint against deliberate data fabrication.

On the other hand, we assume that Bob, the verifier, or other potential verifiers, may be malicious. Bob may attempt to infer sensitive information, such as whether a particular individual’s genomic data was used in Alice’s research, even though no raw data is shared. Such a malicious verifier could exploit the additionally shared data from Alice and auxiliary information from public datasets to launch membership inference attacks (MIAs). The goal of these attacks is to determine if a specific individual, or "victim," is part of the dataset associated with a particular trait or disease, thereby revealing sensitive information about the individual’s health status.

\vspace{-3mm}
\section{Methodology}
\label{section:methodology}

We provide the details of the proposed scheme below.

\vspace{-3mm}
\subsection{Reverse Engineering the Contingency Tables}
\label{section:reverse_engineering}

The core idea behind our approach is the relationship between Minor Allele Frequency (MAF) and the contingency table for a specific SNP $i$. Specifically, for SNP $i$, following the example in Table~\ref{table:contingency_table}, we represent the contingency table as $\textbf{CT}^i = \{T_s^i, C_s^i\}$, where $T_s^i = \{T_0^i, T_1^i, T_2^i\}$ represents the counts for the case group, and $C_s^i = \{C_0^i, C_1^i, C_2^i\}$ represents the counts for the control group. Using the case group as an example, we can calculate the MAF for SNP $i$ in the case group as follows:

\begin{equation}
\text{MAF}_i = \frac{2T_2^i + T_1^i}{2(T_0^i + T_1^i + T_2^i)}
\label{equation:maf}
\end{equation}

This formula relates the contingency table directly to the MAF, allowing us to calculate MAF values for each SNP and for both the case and control groups when the contingency table is known. However, the reverse process—deriving a contingency table from the MAF—is more challenging, as multiple configurations of $T_0^i$, $T_1^i$, and $T_2^i$ could produce the same MAF. This introduces uncertainty when attempting to reverse-engineer the contingency table from MAF data.

\begin{table}[h]
\footnotesize
\centering
\vspace{-3mm}
\begin{tabular}{lccc}
        \hline
        & \multicolumn{3}{c}{Genotype} \\
        & 0 ($\mathbf{AA}$) & 1 ($\mathbf{AG}$) & 2 ($\mathbf{GG}$) \\ \hline
        Case ($T_s$)    & $T_0$  & $T_1$  & $T_2$ \\
        Control ($C_s$) & $C_0$  & $C_1$  & $C_2$ \\
        \hline
\end{tabular}
\vspace{2mm}
\caption{A contingency table $\textbf{CT}$ showing genotype distribution for cases and controls, where the minor allele is $\mathbf{G}$ and the major allele is $\mathbf{A}$. }
\vspace{-8mm}
\label{table:contingency_table}
\end{table}

The key to our approach is that we utilize the p-values reported in GWAS. For SNP $i$, we have the row $C_s^i$ representing the control group in the contingency table, which can be derived from public datasets. Our task is to find the best match for the row $T_s^i$ representing the case group in the same contingency table \textbf{CT}$^i$. Each GWAS experiment generates a significance p-value representing the strength of the association between SNPs and phenotypic traits. We utilize this p-value and perform a \textbf{grid search} to find the best match $T_s^i$ that generates a p-value closest to the reported one.

The grid search enumerates all possible distributions of the case group's genotype, constrained by the number of samples $n_t$ in the case group. This is equivalent to allocating $n_t$ elements into 3 categories, with the number of possible combinations given by $\binom{n_t + 2}{2}$.
We generate all valid combinations of $T_s^{i'}=\{T_0^i, T_1^i, T_2^i\}$ that satisfy $T_0^i + T_1^i + T_2^i = n_t$, where each combination $T_s^{i'}$ represents a potential genotype distribution. For each, we compute the p-value using the same statistical test Alice applied.
Finally, we select the combination whose p-value is closest to the reported study value. If multiple combinations match, we randomly select one as the target $T_s^i$.

Importantly, this reverse-engineering process may lead to discrepancies between the estimated and actual contingency tables. However, this is not a significant issue in our validation process, as we do not require the reverse-engineered contingency table to be an exact match. Instead, we tolerate some deviations and implement a relaxed detection method, as described in the next section.

We repeat this process for each SNP to reverse-engineer the case rows $T_s^i$, which will then be used in the validation procedure described in the next section.

\vspace{-2mm}
\subsection{MAF-Based Validation of GWAS Outcomes}

Once the $T_s^i$ has been reconstructed, we validate the GWAS outcomes by comparing the MAF values derived from the reverse-engineered contingency table with publicly available MAF statistics for the correponding phenotype. The goal of this comparison is to detect significant discrepancies that may indicate errors in the original research. 
To perform this validation, we first calculate the MAF for each SNP for the case group using Equation~\ref{equation:maf} from the case rows $T_s^i$ that are derived from Section~\ref{section:reverse_engineering}. 

Next, we calculate the Hamming distance between the MAF values we have generated and the public MAF values for the corresponding phenotype. The Hamming distance quantifies the difference between the two sets of MAF values for each SNP. We then compute the average Hamming distance across all SNPs as:
\vspace{-3mm}
\begin{equation}
\text{Average Hamming Distance} = \frac{1}{n} \sum_{i=1}^{n} | \text{MAF}_{i,\text{calculated}} - \text{MAF}_{i,\text{public}} |,
\label{equation:avg_hamming_distance}
\end{equation}
where $n$ is the number of SNPs. If the average Hamming distance exceeds a predefined threshold, we flag the GWAS outcomes as suspicious and in need of further investigation. Otherwise, we conclude that the research findings are likely reliable.

It is important to note that we do not expect a perfect match between the calculated and public MAFs due to potential errors in the reverse-engineering of the contingency table and inherent biological variability. Therefore, our validation process does not assume that a zero Hamming distance implies correctness or that any non-zero distance implies an error. Instead, we introduce a tolerance threshold to account for minor deviations. We only consider potential errors in the research outcomes if the average distance exceeds this threshold.

\begin{figure}[ht]
    \centering
    \includegraphics[width=1\linewidth]{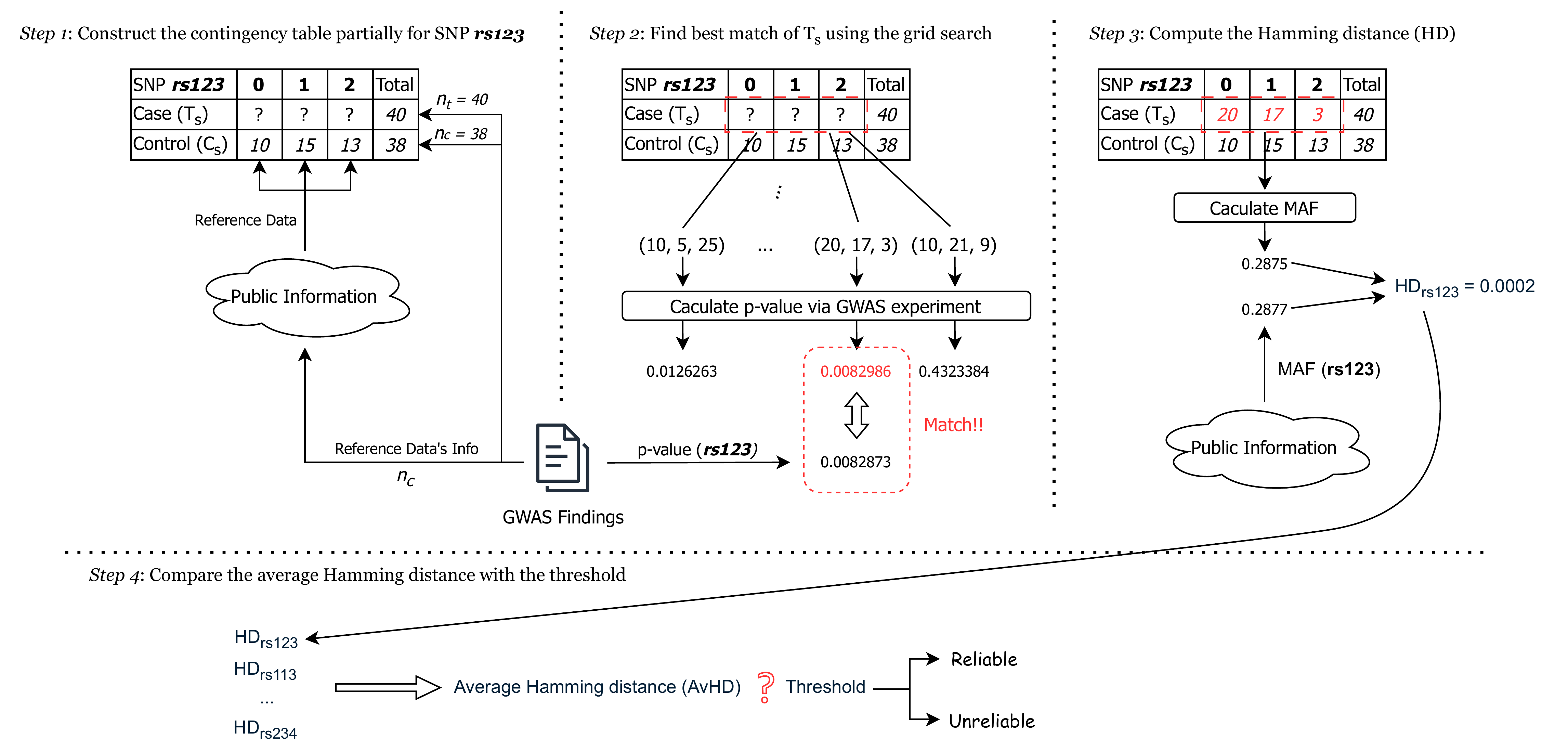}
    \caption{Bob's workflow for validating GWAS outcomes.}
    \label{figure:workflow}
    
    \vspace{0.2cm}
    
    Bob's workflow consists of 4 steps, where the first 3 steps are applied to each SNP:
    \begin{itemize}
        \item \textbf{Step 1:} For each SNP $i$, Bob constructs a contingency table, where the control row $C_s^i$ and the sample sizes are prefilled using public sources and Alice's reported GWAS findings.
        \item \textbf{Step 2:} Bob conducts a grid search over all possible values of the case row $T_s^i$ to find the best match that reproduces the p-value reported by Alice’s GWAS experiment.
        \item \textbf{Step 3:} Bob computes the Hamming distance between the Minor Allele Frequency (MAF) derived from the reconstructed case row $T_s^i$ and the public MAF associated with the target phenotype.
        \item \textbf{Step 4:} After calculating the Hamming distances for all SNPs, Bob compares the average Hamming distance (AvHD) with a predefined threshold to determine the reliability of the GWAS outcomes.
    \end{itemize}
    \vspace{-8mm}
\end{figure}

Figure~\ref{figure:workflow} shows an example of the validation process of Alice's GWAS findings by Bob. The source code is can be accessed online in Github.\footnote{\href{https://github.com/SpidLab/Validating-GWAS-Findings-through-Reverse-Engineering-of-Contingency-Tables}{https://github.com/SpidLab/Validating-GWAS-Findings-through-Reverse-Engineering-of-Contingency-Tables}}

\vspace{-3mm}

\section{Evaluation}
\label{section:evaluation}

\vspace{-2mm}
\subsection{Dataset}
We use three datasets with different phenotypes: lactose intolerance, hair color, and handedness, obtained from the OpenSNP project~\cite{opensnp}. The lactose intolerance dataset contains 9,091 SNPs from 60 individuals. The hair color dataset includes 9,686 SNPs, also from 60 individuals. The handedness dataset consists of 6,985 SNPs from 59 individuals.
\vspace{-2mm}
\subsection{Modeling GWAS Errors}
We model GWAS errors in two primary ways. In the first scenario, we assume that Alice, the researcher, may: 1) record SNP data in the wrong columns, 2) load incorrect columns during data pre-processing, 3) make calculation errors during the GWAS analysis, or 4) report the wrong SNPs when generating the findings. As a result, the findings contain some correct SNPs but also include incorrect ones, where certain correct SNP IDs are replaced with non-significant ones. We model all these types of errors using the \textbf{reporting error}. This error is characterized by an error factor, $\beta_1$, which represents the proportion of incorrect SNPs reported due to these mistakes, while the remaining $1 - \beta_1$ are still correct. A larger $\beta_1$ indicates fewer correct SNPs and, therefore, a higher number of incorrect SNPs in the published findings.

In the second scenario, all results might be influenced by slight noise during calculations, leading to incorrect p-values being reported for all SNPs, even though the noise itself is minimal. We model this type of error in the reported p-values as a \textbf{deviation error}. In this case, the p-values are distorted by the addition of normally distributed noise. This error is characterized by $\beta_2$, which represents the standard deviation of the noise added to the p-values. A larger $\beta_2$ indicates greater distortion in the p-values due to increased noise.

It is important to note that these errors are unintentional and independent of each other, making the simultaneous occurrence of both errors unlikely. Even if both errors were to occur, it would be highly improbable for the combination of errors to produce the same p-values as the original, error-free findings. As a result, we do not model experiments involving both errors together, as the combined effect would still lead to significant deviations. These deviations would make it unlikely for the results to match the publicly available MAFs. If a mismatch in MAF is found, the GWAS findings should be reconsidered and further investigated. For simplicity, we model these two errors separately.
\vspace{-2mm}
\subsection{Experiment Setting}
We assume that Alice uses the $\chi^2$ test in her GWAS analysis, and we replicate this by running chi-square experiments on the three datasets. For each dataset, we create a reference dataset that does not contain the corresponding phenotype and use it to simulate the publicly available reference dataset Alice would have used in her GWAS. We set $\alpha = 0.05$, meaning that all SNPs with a p-value less than 0.05 are considered significantly associated with the phenotype and will be included in the final GWAS findings. For the error factors, we vary $\beta_1$ from 0 to 1, representing 0\% to 100\% of truly significant SNPs included in the GWAS report. For $\beta_2$, we introduce a very small amount of noise, ranging from 0 to 0.01, to test the sensitivity of our detection mechanism. All experiments are repeated 10 times, and we take the average results to ensure reliability. 
\vspace{-2mm}
\subsection{GWAS Outcomes Validation Against Errors}
\begin{figure}[ht]
    \centering\vspace{-5mm}
    \begin{subfigure}[b]{0.45\textwidth}
        \centering
        \includegraphics[width=\textwidth]{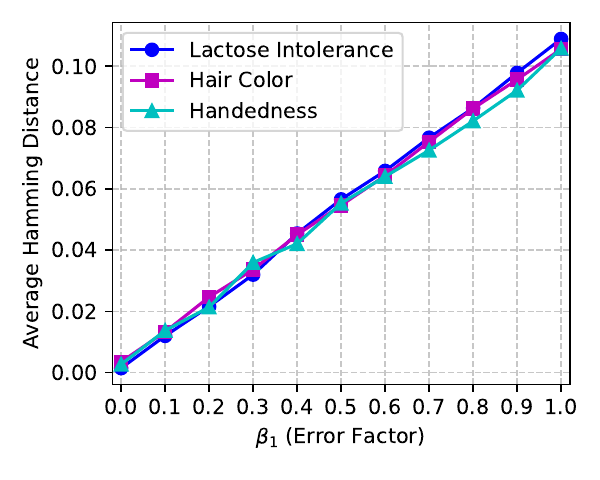}
 
        \caption{Vs. Reporting Error}
        \label{figure:reporting_error}
    \end{subfigure}
    \hfill
    \begin{subfigure}[b]{0.45\textwidth}
        \centering
        \includegraphics[width=\textwidth]{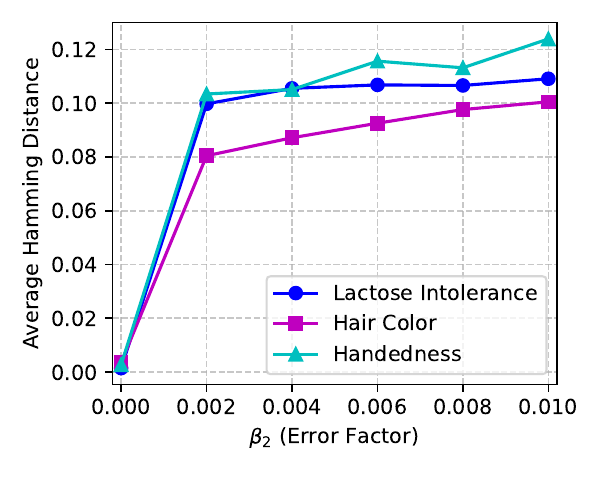}
        \caption{Vs. Deviation Error}
        \label{figure:deviation_error}
    \end{subfigure}
    \caption{Performance of GWAS outcome validation against (a) reporting error and (b) deviation error, evaluated across three datasets.}
    \label{figure:combined_error_vs_avhd}
    \vspace{-5mm}
\end{figure}

We present the impact of the average Hamming distance (AvHD) as a function of the reporting error in Figure~\ref{figure:reporting_error}. As shown in the figure, our scheme exhibits a consistent increase in the AvHD as the error factor grows. In the absence of error, the AvHD values are 0.0014, 0.03, and 0.02 for the lactose intolerance, hair color, and handedness datasets, respectively. However, when Alice incorrectly reports only 5\% of significant SNPs, our scheme detects an increase in AvHD to 0.011, 0.013, and 0.013 for the same datasets, with a generally linear upward trend. This demonstrates that even a slight error leads to a noticeable change in AvHD, allowing Bob to identify potential errors.

For deviation error, we present the results in Figure~\ref{figure:deviation_error}. A clear difference emerges between the error-free scenario and the presence of deviation error. Even with a small deviation factor of $\beta_2 = 0.002$, the AvHD quickly rises above 0.08 across all three datasets, indicating a sharp increase. This noticable jump makes it evident that an error has occurred in the GWAS experiment. These results demonstrate that our scheme is also effective at detecting deviation errors.

Note that although we observe clear distinctions in AvHD between error and error-free results, we do not claim to have established a single strict threshold. The focus of this work is not on precisely determining an exact threshold, but rather on demonstrating that such a threshold can be easily identified. This is possible because our proposed method consistently generates a noticeable difference between correct and erroneous outcomes.

\subsection{Impact of Precision in Shared p-values}
Alice may not always share highly accurate p-values with the GWAS findings. In some cases, the p-values associated with the reported SNPs can be pruned, which can affect the validation performance. To explore this, we conducted an experiment to assess how varying precision impacts our scheme's performance. As shown in Figure~\ref{figure:precision_impact} for the lactose intolerance dataset, the average Hamming distance (AvHD) remains stable until the number of decimal places is reduced to 6. This indicates that as long as Alice provides p-values with at least 7 decimal places (or even 6), our scheme performs similarly to using maximum precision (which is 20 decimal places in our experiments). High precision in p-value reporting ensures the most reliable validation, as maintaining 7 or more decimal places maximizes the scheme’s ability to consistently and accurately detect errors, offering stronger confidence in GWAS outcome verification.

It's also important to note that while AvHD increases as the precision of the p-values decreases, this does not mean that our scheme loses its effectiveness at lower precision levels. For example, even with p-values rounded to 3 decimal places, a clear rise in AvHD can still be observed as errors increase based on our experiments. This demonstrates that our scheme can reliably detect errors even when the precision of the p-values is reduced, maintaining robust performance across various levels of precision.

\vspace{-3mm}

\subsection{Impact of MAF Deviations Between Private and Public Datasets}

\begin{figure}[ht]
    \vspace{-5mm}
    \centering
    \begin{subfigure}[b]{0.45\textwidth}
        \centering
        \includegraphics[width=\textwidth]{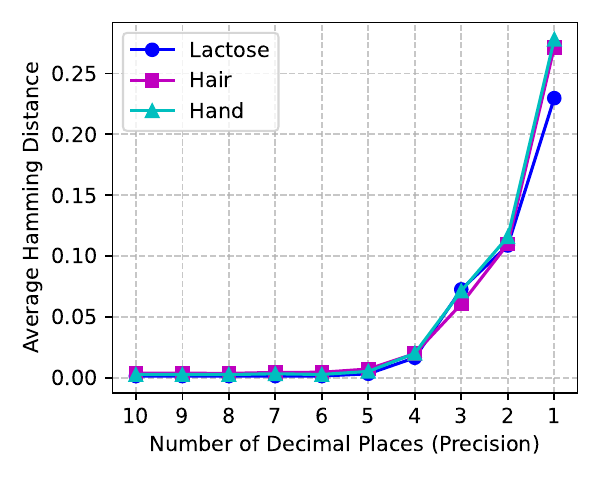}
        \caption{Impact of p-value precision}
        \label{figure:precision_impact}
    \end{subfigure}
    \hfill
    \begin{subfigure}[b]{0.45\textwidth}
        \centering
        \includegraphics[width=\textwidth]{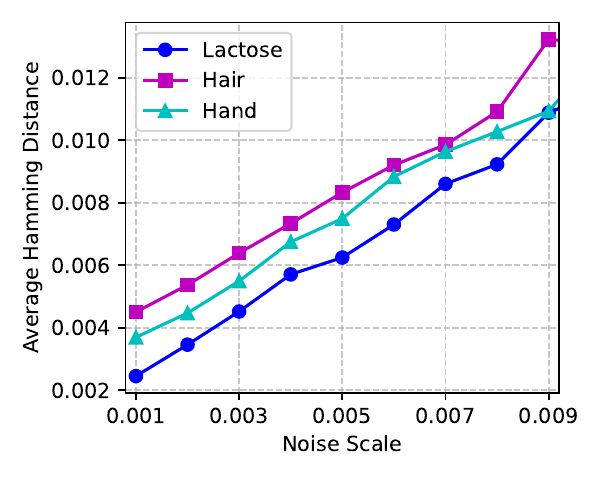}
        \caption{Impact of deviated MAFs}
        \label{figure:deviated_maf_impact}
    \end{subfigure}
    \caption{Performance of GWAS outcome validation against (a) varying p-value precision and (b) MAF deviations, evaluated across three datasets.}
    \label{figure:impact}
\vspace{-7mm}
\end{figure}

We use public MAFs associated with the same trait as in the target dataset for error detection. However, it is possible that the MAFs in the target dataset and the public dataset may deviate slightly. To evaluate this impact, we model this scenario by adding Laplacian noise to assess whether such deviations affect the GWAS outcome validation results. As shown in Figure~\ref{figure:deviated_maf_impact}, the average Hamming distance (AvHD) remains small under low noise levels, indicating that the same threshold used for non-deviated MAFs can still be applied. However, the AvHD increases linearly with the noise scale. Therefore, it is important that the two MAFs are not too far apart to ensure optimal detection performance.

\vspace{-5mm}
\section{Discussion}
\label{discussion}
As discussed earlier, the proposed mechanism can also be utilized by Alice for a "self-check" to validate her own GWAS findings before publication. While our work primarily focuses on the researcher-verifier setting, this approach is equally beneficial for researchers who wish to confirm the accuracy of their results independently.

Specifically, Alice can perform a local validation on her GWAS findings by following the same steps as Bob. She would input the GWAS findings, public reference data, and the sample sizes of both the case and control groups. Then, by applying the validation method described in Section~\ref{section:methodology}, Alice can compare the MAFs calculated from her contingency tables against the public MAFs. This allows her to verify her results prior to publication.

This built-in self-validation mechanism not only helps Alice ensure the accuracy of her findings but it also allows her to confidently state that the proposed validation steps have been completed. By doing so, she can provide greater assurance to the research community, strengthening the credibility of her GWAS findings.

\vspace{-1mm}

\section{Conclusion}

In this paper, we proposed a novel method for validating GWAS findings without sharing sensitive genomic data, relying on public reference data and reported p-values to reverse-engineer contingency tables. By comparing MAFs from these tables with public MAFs, our approach effectively identifies unintentional errors throughout the GWAS process, providing reliable validation. The method performs well even with reduced precision in the reported p-values, ensuring robustness. This approach enhances trust in genomic research by offering a practical solution for reproducibility validation in genomic studies.

\bibliographystyle{splncs04}
\bibliography{bib}
\section{Appendix}
\appendix
\section{Background}
\label{section:background}

\subsection{Single Nucleotide Polymorphism (SNP)}
Single nucleotide polymorphisms (SNPs) are the most common type of genetic variation among individuals, involving differences at a single nucleotide position in DNA sequences.  SNPs can significantly correlate with traits and diseases, making them valuable for understanding disease susceptibility and individual responses to medications or environmental factors. In genome-wide association studies (GWAS), SNPs serve as markers to investigate potential links between genetic variations and traits or diseases.

\subsection{Minor Allele Frequency (MAF)}
In the context of a biallelic SNP, where two possible allele variants exist at a single position, minor allele frequency (MAF) refers to the frequency of the less common allele in a population. MAF is essential for determining the significance of an SNP in association studies and is frequently used in GWAS to evaluate differences in genetic variants between case and control groups.

\subsection{Contingency Table}
A contingency table generally contains the counts of individuals with and without a particular phenotype, specifically a case group (individuals exhibiting a specific phenotype or trait) and a control group (individuals without the phenotype or trait). In our context, we further categorize these counts based on SNP values within both the case and control groups. For example, Table~\ref{table:contingency_table} shows the distribution of individuals in both groups with specific genotypes, which facilitates statistical analysis of the observed data.

\end{document}